\documentclass[twocolumn,groupedaddress,
superscriptaddress,preprintnumbers,
amsmath,amssymb,
aps,
]{revtex4}
\usepackage{graphicx}
\usepackage{dcolumn}
\usepackage{bm}%
\usepackage[colorlinks=true,citecolor=blue]{hyperref}
\DeclareUnicodeCharacter{2032}{\ensuremath{'}}
\hypersetup{colorlinks=true,citecolor=blue,linkcolor=red,urlcolor=blue}

\usepackage{float}
\usepackage{amsmath}

\usepackage{bm}
\newcommand{\beq}{\begin{equation}}
\newcommand{\eeq}{\end{equation}}
\newcommand{\beqnar}{\begin{eqnarray}}
\newcommand{\eeqnar}{\end{eqnarray}}
\newcommand{\bfig}{\begin{figure}}
\newcommand{\efig}{\end{figure}}

\usepackage{color}
\usepackage[dvipsnames]{xcolor} 

\begin{document}

\title{Enhanced Brewster Angle Shift in Doped Graphene via the Fizeau Drag Effect}

\author{Rafi Ud Din}
\affiliation{Department of Physics, Zhejiang Normal University, Jinhua, Zhejiang 321004, China}
\affiliation{Zhejiang Institute of Photoelectronics \& Zhejiang Institute for Advanced Light Source, Zhejiang Normal University, Jinhua, Zhejiang 321004, China.}
\author{Yuncheng Zhou}
\affiliation{Huazhong Institute of electro-optics, Wuhan, Hubei 430000, China.}
\author{Reza Asgari}
\email{asgari@ipm.ir}
\affiliation{Department of Physics, Zhejiang Normal University, Jinhua, Zhejiang 321004, China}
\affiliation{School of Physics, Institute for Research in Fundamental Sciences (IPM), Tehran 19395-5531, Iran}
\author{Gao Xianlong}
\email{gaoxl@zjnu.edu.cn}
\affiliation{Department of Physics, Zhejiang Normal University, Jinhua, Zhejiang 321004, China}


\date{\today}

\begin{abstract}
We derive the general Fresnel coefficients for reflection by incorporating the Fizeau drag effect in doped graphene, which arises from the unique behavior of its massless Dirac electrons. Using the standard Maxwell equations and constitutive relations, we analyze the influence of this relativistic phenomenon on the optical properties of doped graphene. Our study focuses on the angular shift of Brewster's angle in a structure where monolayer graphene is sandwiched between two static dielectric media.
Our findings reveal that the presence of the Fizeau drag effect significantly enhances the Brewster angle shift, leading to substantial modifications in the optical characteristics of the graphene channel, including notable alterations in the reflectance spectrum. We demonstrate that this angular shift can be further amplified by increasing the drift velocities and charge densities of the electrons in graphene, offering a tunable mechanism for controlling optical behavior in graphene-based systems. The findings of this work have significant implications for the design and development of planar photonic devices that take advantage of the optical characteristics of graphene. This breakthrough creates new opportunities for the use of graphene in sophisticated photonic technologies, where exact control over the interactions between light and matter is essential.
\end{abstract}

\maketitle


\section{Introduction}

The Fresnel coefficients, which are obtained from the Maxwell equations, constitutive relations, and boundary conditions, provide the reflection and transmission properties of light impinging on the interface between two distinct media~\cite{jackson2021classical}. This process is also used for interfaces between unusual media, such as those that separate bi-isotropic dielectrics \cite{sihvola1992properties} and topological insulators, for which the chiral coupling magnitude may be conveniently determined by polarization rotation at the Brewster angle (BA) \cite{saleh2019fundamentals}, which is the angle of incidence at which light with a particular polarization is perfectly transmitted through a transparent material with no reflection, and at this angle, the reflected and refracted rays are perpendicular to each other. 

The BA  was first observed by Brewster more than two centuries ago~\cite{doi:10.1098/rstl.1815.0010} and was later confirmed by Fresnel~\cite{Buchwald1991TheRO}. At Brewster’s angle, an unpolarized light becomes linearly polarized. Due to its wide applications in polarization filters~\cite{Lv:12}, material sensors~\cite{Valiyaveedu2019}, and microscopy~\cite{Stuart1996}, BA has recently been revisited in media with electron sea such as dense plasma~\cite{PhysRevE.94.043202}, two-dimensional (2D) materials~\cite{PhysRevA.109.L031502}, superlattice structures~\cite{PhysRevB.61.12877}, metamaterials~\cite{PhysRevB.73.193104} as well as for surface plasmons~\cite{PhysRevLett.106.123902}. However, manipulation of these effects demands special attention and efficient strategies in fast-moving electron media. Among these materials, graphene is attracting tremendous interest for photoelectronic applications due to its unique properties~\cite{novoselov2012roadmap,PhysRevB.87.121402}.

It has emerged as an ideal material due to its tunable conductivity through electrostatic gating~\cite{blake2007making}. Previously, the BA is controlled in graphene by varying its surface conductivity through an applied voltage~\cite{Zefeng2018,Majerus_2018}. The mechanism was also used to develop a graphene/quartz modulator~\cite{Zefeng2018}. The valley dependence of BA in strained graphene has also been reported~\cite{Wu2011valley}. The Brewster angle was shown to undergo an up-shift upon the insertion of 2D materials in dielectric interfaces~\cite{PhysRevA.109.L031502,Majerus_2018}. However, the relativistic effects arising from passing a direct current through graphene, leading to Doppler shifts, are not yet studied and are our concerns in this paper.

The drag effect leads to slight modification in the light's speed when it is propagating through moving media. This effect was first discovered by Fizeau in 1851~\cite{Fizeau} and was verified in the flow-water tube experiment~\cite{PhysRevResearch.4.033124,PhysRevA.86.013806,Kuan2016}. 
This Fizeau drag effect describes how the speed of light is affected when it travels through a fluid (like water) that is itself moving relative to the observer. The light's velocity difference along the two directions leads to a shift in the interference fringes.

The high electron mobility in graphene~\cite{geim2007rise,geim2009graphene} and the propagation of its massless Dirac fermions have been used experimentally to demonstrate the Fizeau drag effect for graphene plasmons~\cite{Dong2021Fizeau,Zhao2021efficient}. The effect was observed as a Doppler shift in wavelength of plasmonic modes effectively dragged by Dirac electrons along their direction of propagation. The idea of plasmon Fizeau drag by fast-moving Dirac electrons has also been extended to three-dimensional materials~\cite{doi:10.1021/acsphotonics.3c01416}. The Goos-H\"{a}nchen shifts in reflection for a light beam within a graphene structure influenced by the Fizeau drag effect have recently been explored~\cite{PhysRevB.109.115403}.

In this paper, motivated by recent findings on the BA~\cite{PhysRevA.109.L031502}, we explore modifications to the BA in graphene at zero temperature by incorporating relativistic effects induced by its fast-moving, massless Dirac electrons. We demonstrate that the BA shift can be enhanced by several degrees due to the Fizeau drag effect on the incoming wave. Furthermore, the BA shift can be potentially controlled by adjusting the drift velocity and the number of charged particles in graphene. We have modified the expressions for conductivity, reflection coefficient, and BA shift to account for the Fizeau drag in graphene. These analytical expressions are supported by numerical results. Our study presents an efficient platform for guiding light through current-carrying graphene. 

The paper is organized as follows. In Sect. \ref{sec:Int}
, we begin by discussing the model and calculating the Fresnel optical coefficients for a doped 2D material positioned between two dielectric media. Next, we explore the Fizeau drag effect and extend the reflection and transmission analysis accordingly, including the conditions under which the BA can be observed. In Sect. \ref{sec:Num}, we present and discuss our numerical calculations. Finally, we summarize our key findings in Sec. \ref{sec:Con}.

\section{Model and Theory}\label{sec:Int}
\begin{figure}[t]
\centering
\includegraphics[width=3.4in]{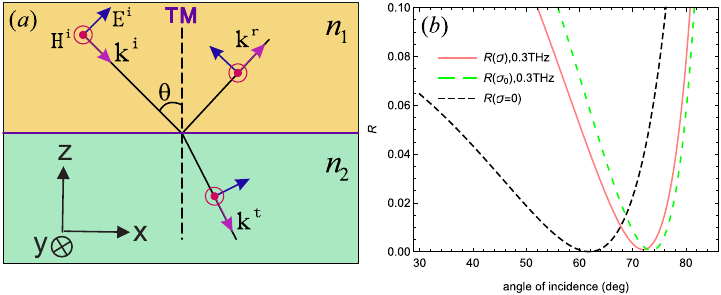}
\caption{(Color online) ($a$) Geometry for analyzing the BA shift for a TM-polarized plane wave incident from medium 1 onto medium 2. $n_1$ ($n_2$) denotes the refractive index of medium 1 (2) and $\theta$ is the angle of incidence of the wave. ($b$) Reflectance angular spectrum of TM-polarized wave as a function of incident angle for three different cases. The black dotted line denotes the case of no graphene layer between the two media ($\theta^i_{B}$-case), the pink solid line represents a normal graphene channel of conductivity $\sigma$ sandwiched between the two media ($\theta_B$-case), while the green curve is for the graphene of conductivity $\sigma_0$ under the effect of Fizeau drag ($\theta^0_{B}$-case).}\label{1}
\end{figure}

We consider a doped graphene with finite conductivity $\sigma$, encapsulated between two static dielectric media at zero temperature. The geometry depicted in Fig.~\ref{1}(a) illustrates our model for the investigation of the BA shift for a transverse magnetic (TM)-polarized plane wave under the influence of the Fizeau drag effect in doped graphene. A TM-polarized plane wave is incident from medium 1 of the refractive index $n_1$ to medium 2 of the refractive index $n_2$ in the $xz$ plane with dielectric constant $\varepsilon_1$ and $\varepsilon_2$, respectively. The incident, reflected, and transmitted wave vectors are represented as $\textbf{k}^i$, $\textbf{k}^r$, and $\textbf{k}^t$, respectively. $\theta$ is the angle of incidence. The usual boundary conditions for the electromagnetic fields in the geometry shown in Fig.~\ref{1}(a) are
\begin{eqnarray}
&&E_{2x}-E_{1x}=0,\nonumber\\
&&H_{2y}-H_{2y}={\sigma} E_{2x}.\label{bc1}
\end{eqnarray}
But $E_{1x}=E^i_{x}+E^r_{x}$ and $E_{2x}=E^t_{x}$, and using similar equations for $H$, we get
\begin{eqnarray}
&&E^i_{x}+E^r_{x}=E^t_{x},\nonumber\\
&&H^t_{y}-(H^i_{y}+H^t_{y})= {\sigma} E^t_{x}.\label{bc2}
\end{eqnarray}
Now using the Maxwell's equation $\nabla\times \textbf{H}=\dfrac{\partial \textbf{D}}{\partial t}$ in medium 1 and 2, we have
\begin{eqnarray}
\frac{\partial}{\partial z} (H^i_y+H^r_y)&&=ik \varepsilon_1 (E^i_x+E^r_x) ,\nonumber\\
\frac{\partial}{\partial z} H^t_y&&=ik \varepsilon_2 E^t_x .
\end{eqnarray}
These equations further reduced to
\begin{eqnarray}
k^i_z H^i_y-k^r_z H^r_y&&=k \varepsilon_1 (E^i_x+E^r_x) ,\nonumber\\
k^t_z H^t_y&&=k \varepsilon_2 E^t_x .
\end{eqnarray} 
From the above expressions, the reflection ($r$) and transmission ($t$) coefficients are obtained via $r=H^r_y/H^i_y$ and $t=H^t_y/H^i_y$ as,
\begin{equation}
r=\frac{\varepsilon_2/k^t_z-\varepsilon_i/k^i_z+\sigma/k}{\varepsilon_2/k^t_z+\varepsilon_i/k^i_z+\sigma/k},\label{ref1}
\end{equation}
\begin{equation}
t=\frac{2\varepsilon_2/k^t_z}{\varepsilon_2/k^t_z+\varepsilon_i/k^i_z+\sigma/k}.\label{trans1}
\end{equation}
Here, $k^i_z=n_1k\cos\theta$, $k^t_z=n_1k\sqrt{n^2-\sin^2\theta}$ and $k_x=n_1k\sin\theta$ such that $k=\omega/c$ is the wave vector in vacuum with $\omega$ its frequency. In the preceding relations, $n_1=\sqrt{\mu_1\varepsilon_1}$ and $n_2=\sqrt{\mu_2\varepsilon_2}$ are the refractive indices of the lower and upper medium, respectively, and $\mu_i$ is the permeability of the $i$th medium. In the case of nonmagnetic media, the above expressions are reduced to
\begin{equation}
r=\frac{\varepsilon_2+\sigma\sqrt{\varepsilon_1(\varepsilon-\sin^2\theta})-\varepsilon_1\sec\theta\sqrt{\varepsilon-\sin^2\theta}}{\varepsilon_2+\sigma\sqrt{\varepsilon_1(\varepsilon-\sin^2\theta})+\varepsilon_1\sec\theta\sqrt{\varepsilon-\sin^2\theta}},\label{ref}
\end{equation}
\begin{equation}
t=\frac{2\varepsilon_2}{\varepsilon_2+\sigma\sqrt{\varepsilon_1(\varepsilon-\sin^2\theta})+\varepsilon_1\sec\theta\sqrt{\varepsilon-\sin^2\theta}},\label{trans}
\end{equation}
where $\varepsilon=\varepsilon_2/\varepsilon_1$ and $\sigma(\omega)$ is the complex conductivity of graphene. The reflectance can be obtained from $R=|r|^2$. It is worth noting that the Fresnel coefficients can be easily generalized for cases where the environment is a magnetic material ~\cite{saleh2019fundamentals}.

The conductivity of doped graphene at zero temperature, which can be derived from the Kubo formula \cite{gusynin2006magneto} or the random phase approximation \cite{platzman1973waves}, is: 
\begin{eqnarray}\label{conduc1}
\sigma(\omega)=&&\frac{e^2\epsilon_F}{\pi\hbar^2}\frac{i}{\omega+i\gamma}\nonumber\\
&&+\frac{e^2}{4\hbar}\bigg\{\vartheta(\hbar\omega-2\epsilon_F)+\frac{i}{\pi}\ln\left|\frac{\hbar\omega-2\epsilon_F}{\hbar\omega+2\epsilon_F}\right|\bigg\}.
\end{eqnarray} 
Here, $\epsilon_F = \hbar v_F k_F$ represents the Fermi energy, where $k_F = \sqrt{\pi \rho}$ is the Fermi wave vector, and $v_F = 1 \times 10^6$ m/s is the Fermi velocity of the charged particles. The parameter $\gamma$ denotes the scattering rate of the charge carriers with density $\rho$, and $\vartheta$ is the step function. The first term in the expression above accounts for intraband electron-photon scattering, while the second term arises from interband electron transitions~\cite{hanson2008quasi}. It should be noted that this expression applies to highly doped or gated graphene. In the long-wavelength and high-doping limit, where $\omega \ll v_F k_F$, the interband contributions become negligible, and the intraband term dominates, defining the total conductivity of graphene.

We assume that the Fizeau drag that affects the incident light is solely a result of the drifting electrons in graphene. These drifting electrons are considered to propagate along the positive $x$-axis. To analyze the BA shift in the current-carrying state, we begin with Lorentz transformations applied to various physical quantities of graphene~\cite{borgnia2015quasi,Dong2021Fizeau}
\begin{equation}
\omega_0=\Gamma(\omega-v_D k_x),\label{LT1}
\end{equation}
\begin{equation}
 k_{x,0}=\Gamma(k_x-\frac{v_D}{c^2}\omega),\label{LT2}
\end{equation}
\begin{equation}
 \rho_{0}=\frac{1}{\Gamma}\rho~,\label{LT3}
\end{equation}
where $\omega$ represents the frequency of the incident light, $k_x$ is the horizontal component of the light's wave vector, $\rho$ is the charge density, and $v_D$ denotes the drift velocity of the charged particles in graphene. The subscript ``0" labels quantities measured in the frame moving with velocity $v_D$, and $\Gamma=(1-v_D^2/v_F^2)^{-1/2}$ denotes the Lorentz factor~\cite{Dong2021Fizeau}. By substituting Eqs. (\ref{LT1}), (\ref{LT2}), and (\ref{LT3}) into Eq. (\ref{conduc1}), the conductivity is given by
\begin{eqnarray}
\sigma_0(\omega_0)=\frac{ie^2v_F\sqrt{\pi\rho/\Gamma}}{\pi\hbar\{i\gamma+\Gamma(\omega-v_D k_x)\}}~.\label{conduc2}
\end{eqnarray}
Note that the reflection coefficient in the case of a current-carrying graphene channel can be obtained by inserting the above expression in Eq. (\ref{ref}). As shown, the conductivity depends on the charge density, frequency, and the moving frame velocity $v_D$. Therefore, the BA shift would also depend on these factors as well as the environment refractive indexes. 

The BA condition in the presence of a doped graphene sheet can be obtained from Eq. (\ref{ref}) as
\begin{equation}
 \frac{\varepsilon_2}{k^t_z}-\frac{\varepsilon_i}{k^i_z}+\frac{\sigma}{k}=0.\label{BACondition}
\end{equation}

Solving this equation further (see the appendix for derivation), we get the following quartic equation~\cite{PhysRevA.109.L031502}:
\begin{equation}
 \frac{\sigma'^2}{\eta}x^4-\frac{2\sigma'}{\eta}x^3+[\sigma'^2-n^2-1]x^2-2\sigma'x+1=0,\label{quartic}
\end{equation}
where $\sigma'=\sigma/\sqrt{\varepsilon_1}$, $x=\cos\theta_B$, and $\eta=n^2-1$ such that $n=\sqrt{\mu_2\varepsilon_2}/\sqrt{\mu_1\varepsilon_1}$. We assume that the initial BA at the interface between medium 1 and 2 is $\theta^i_{B}$. By sandwiching a graphene layer between the two media, the modified BA is $\theta_B$ that can be obtained by solving Eq. (\ref{quartic}) numerically. We further assume that the BA under the effect of Fizeau drag in graphene is $\theta^0_{B}$, which can be obtained by replacing $\sigma$ with $\sigma_0$ in Eq. (\ref{quartic}). The shifts of the BA with a graphene layer without and with the Fizeau drag effect are obtained as:
\begin{equation}
\Delta\theta_B(\varepsilon_1;\varepsilon_2;\sigma)=\theta_B(\varepsilon_1;\varepsilon_2;\sigma)-\theta^i_{B}(\varepsilon_1;\varepsilon_2), \label{shift1}
\end{equation}
\begin{equation}
\Delta\theta^0_{B}(\varepsilon_1;\varepsilon_2;\sigma_0)=\theta^0_{B}(\varepsilon_1;\varepsilon_2;\sigma_0)-\theta^i_{B}(\varepsilon_1;\varepsilon_2), \label{shift2}
\end{equation}
and the shift of the BA between the drag-free and drag-affected graphene sheet is:
\begin{equation}
\delta\theta_B=\Delta\theta_{B0}(\varepsilon_1;\varepsilon_2;\sigma_0)-\Delta\theta_B(\varepsilon_1;\varepsilon_2;\sigma). \label{shift3}
\end{equation}

\begin{figure}
\centering
\includegraphics[width=3.2in]{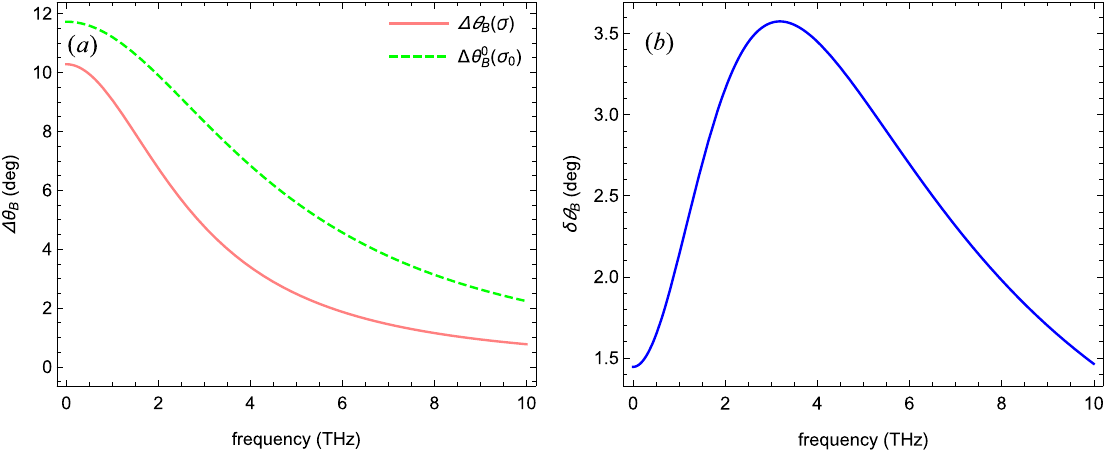}
\caption{(Color online) Shift of the BA as a function of incident frequency. ($a$) BA shift without [pink (solid) curve] and with [green (dashed) curve] Fizeau drag in graphene. ($b$) Difference between the BA shifts when the graphene sheet is without and under the Fizeau drag effect. Here $v_D=0.6v_F$ and $\theta=\theta^i_{B}=61.86^\circ$.}\label{5}
\end{figure}
\begin{figure}[t]
\centering
\includegraphics[width=3.3in]{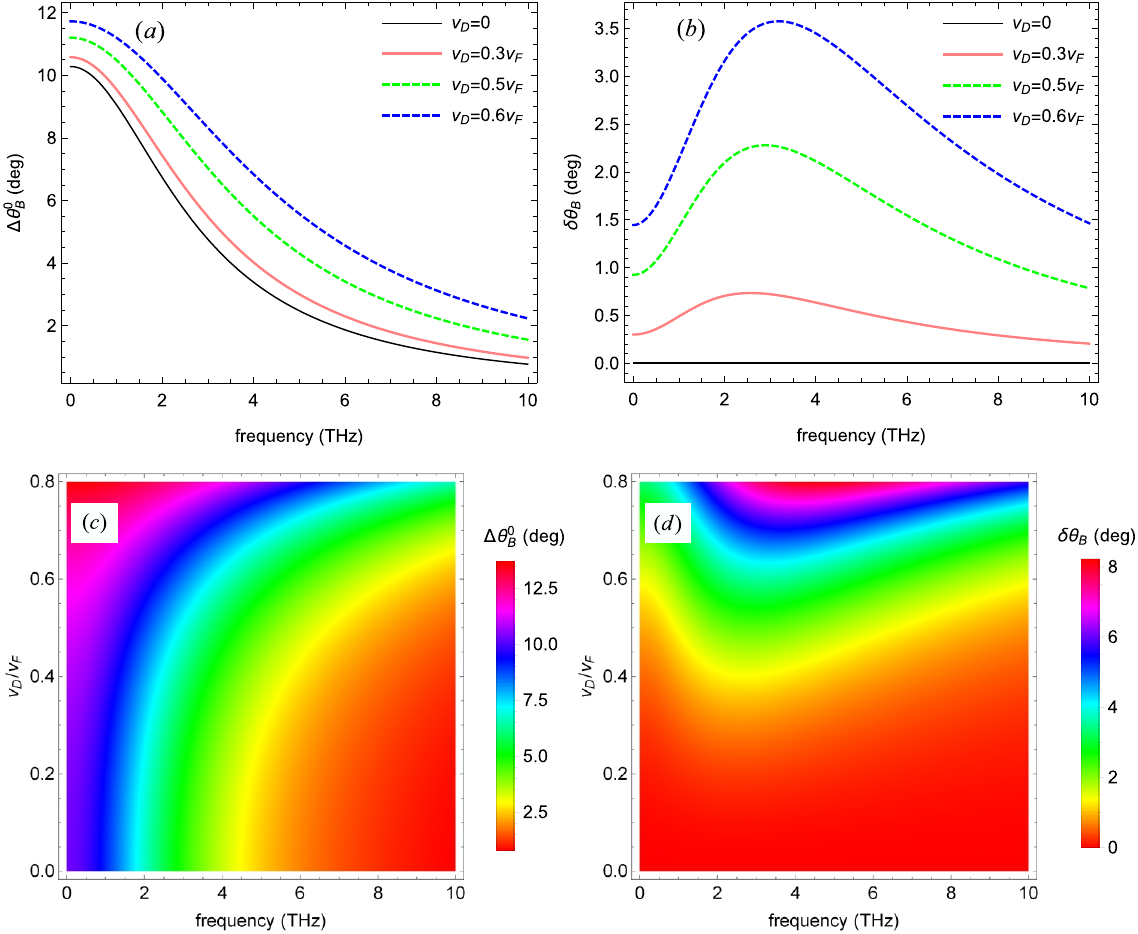}
\caption{(Color online) ($a$) Dependence on the incident frequency of the BA shift when the graphene sheet is under the Fizeau drag effect. ($b$) Difference between the BA shifts when the graphene sheet is without and under Fizeau drag effect as a function of incident frequency. The four lines in each figure correspond to different velocities of the charged particles in graphene. ($c$) Variations of drag-affected BA shift on the frequency and the drifting velocities of the charged particles. ($d$) Difference between drag-free and drag-affected BA shifts as a function of frequency and the drifting velocities of the charged particles. Here $\theta=\theta^i_{B}=61.86^\circ$ and $\rho=5\times10^{12}$cm$^{-2}$.}\label{6}
\end{figure}

\section{Numerical Results and Discussions}\label{sec:Num}
For numerical results, we assume that the carrier density in graphene is $\rho=5\times10^{12}$ cm$^{-2}$ and the scattering rate $\gamma$= 7 meV. Medium 1 is assumed to be air with the relative dielectric constant $\varepsilon_1=1$ and medium 2 is SiO$_2$ substrate with $\varepsilon_2=3.5$ or otherwise stated. In this case, the initial BA is $\theta^i_{B}$=61.86$^\circ$. 

To demonstrate the angular shift of the BA, we first calculate the reflectance spectra of three different cases, that we consider in this study, in Fig.~\ref{1}(b). (1) The case of no graphene layer between the two media is depicted by the black curve, where the initial BA is $\theta^i_{B}$=61.86$^\circ$. (2) A normal graphene channel of conductivity $\sigma$ is sandwiched between the two media. In this case, the BA is shifted to $\theta_{B}$=72$^\circ$ at a frequency of 0.3 THz (Pink dashed curve). (3) The graphene channel is under the Fizeau drag effect. In this case, the minimum reflactance or modified BA is given by $\theta^0_{B}$=73.44$^\circ$ at a speed $v_D=0.6v_F$ of the Dirac electrons at a frequency of 0.3 THz. This shows that, at the given parameters, the BA is shifted towards higher incident angles when Fizeau drag is induced in graphene. The induced Fizeau drag leads to modifications in the optical characteristics of the graphene channel, due to which the reflectance spectrum can be greatly modified.

The observed phenomena can be explained through the understanding that Fizeau drag, as investigated in our study, is a relativistic phenomenon intricately linked to the relative motion of the medium. If we set $v_D$ equal to zero, no drag effects on light are evident, reducing the analysis to a standard graphene channel.

\begin{figure}[t]
\centering
\includegraphics[width=3.3in]{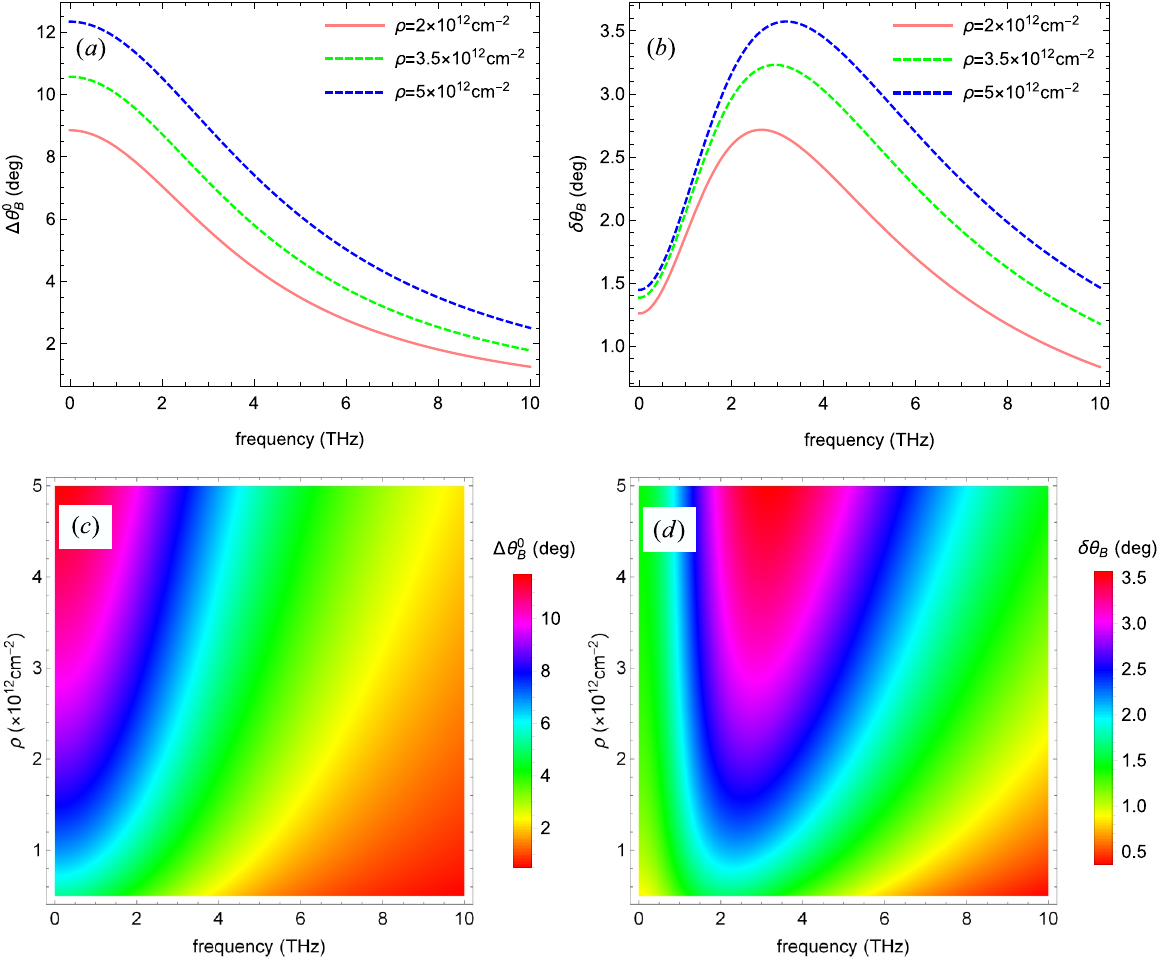}
\caption{(Color online)  ($a$) BA shift in graphene under the influence of Fizeau drag as a function of incident frequency. ($b$) Difference of the BA shifts in graphene with and without Fizeau drag effect. The three lines in each figure correspond to different densities of charged particles in graphene. ($c$) The dependence of drag-affected BA shifts on the frequency and the density of charged particles. ($d$) Difference between drag-free and drag-affected BA shifts as a function of frequency and the density of charged particles. Here $v_D=0.6v_F$ and $\theta=\theta^i_{B}=61.86^\circ$.}\label{7}
\end{figure}

\begin{figure}[t]
\centering
\includegraphics[width=3.3in]{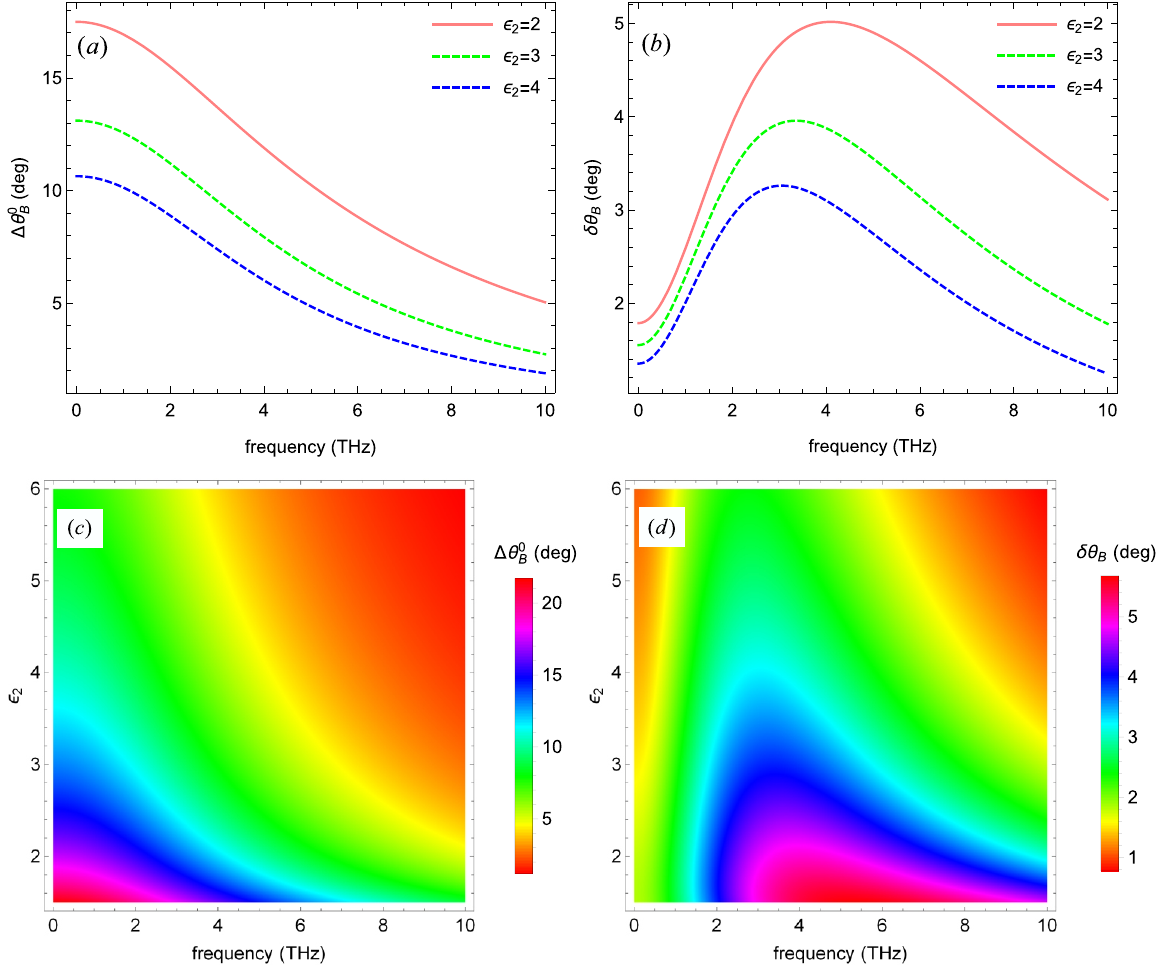}
\caption{(Color online) ($a$) Dependence on the incident frequency of the BA shift when the graphene sheet is under the Fizeau drag effect. ($b$) Difference between the BA shifts when the graphene sheet is without and under Fizeau drag effect as a function of incident frequency. The three lines in each figure correspond to different substrates for graphene. ($c$) The dependence of drag-affected BA shifts on the frequency and the substrate. ($d$) Difference between drag-free and drag-affected BA shifts as a function of frequency and substrate. Here $v_D=0.6v_F$ and $\theta=\theta^i_{B}=61.86^\circ$.}\label{8}
\end{figure}

The BA shifts from the initial value according to Eqs.~(\ref{shift1}) and (\ref{shift2}) are calculated by numerically solving Eq.~(\ref{quartic}) for $\theta_{B}$ and $\theta^0_{B}$. The results obtained are shown in Fig.~\ref{5}, where the solid pink line corresponds to the drag-free graphene case while the green dashed curve is for the drag-affected shift. One can see that the BA shift is higher for intermediate frequencies and decreases smoothly for higher frequencies. The difference between the drag-free and drag-affected BA shifts is higher in the vicinity where the real and imaginary parts of the surface conductivity become equal. It is noteworthy that the induced Fizeau drag leads to BA shift of more than 3.5$^\circ$ even at moderate speeds of the drifting electrons in graphene.

We show how the BA shift is affected when the drifting velocity of the charged particles is increased. This is shown in Fig.~\ref{6}($a$) for different $v_D$ values, where one can notice that the shift increases with increasing drifting velocity. Notice that at $v_D$ = 0 [black curve], the pink curve in Fig.~\ref{5}($a$) is recovered. Likewise, the difference between the drag-free and drag-affected BA shifts also increases with increasing $v_D$ as shown in Fig.~\ref{6}($b$). The dependence of the BA shift in both cases and their difference is also shown with incident frequency and drifting velocity of the electrons in Figs.~\ref{6}($c$) and ($d$), respectively. It is clear from the figure that the BA shift under the effect of the Fizeau drag in graphene $\Delta\theta^0_{B}$ is higher at larger $v_D$ values and low frequencies. However, the difference between BA shifts in the normal and drag-affected graphene channel $\delta\theta_B$ is maximum at intermediate frequencies and higher $v_D$s.

Next, we examine the dependence of the BA shift on the Fermi energy or the density of charged particles in graphene, as illustrated in Fig. \ref{7}. It can be observed that the BA shift in the drag-affected graphene channel increases smoothly with the number of charged particles, as shown in Fig.~\ref{7}($a$) for three different charge density values. Similarly, the difference between the drag-free and drag-affected BA shifts also grows with increasing particle density, as shown in Fig.~\ref{7}($b$). This trend is further clarified in Figs.~\ref{7}($c$) and ($d$), where the dependence of the BA shift and the difference between the shifts in the two cases are plotted against the incident frequency and the number of charged particles. Our findings indicate that the BA shift under the influence of the Fizeau drag in graphene, denoted $\Delta\theta^0_{B}$, is more pronounced at higher densities and lower frequencies. However, the difference between BA shifts in normal and drag-affected graphene channels, denoted as $\delta\theta_B$, reaches its maximum at intermediate frequencies and higher densities.

Finally, we present the substrate dependence of the BA shift in Fig. \ref{8}. As shown in Fig. \ref{8}($a$), the BA shift is more pronounced for substrates with a low dielectric constant (\(\varepsilon_2\)) and decreases for substrates with a high dielectric constant. Similarly, the difference between the drag-free and drag-affected BA shifts is greater for substrates with a low dielectric constant. The dependence of the BA shift in both cases, as well as their difference, is also illustrated with respect to incident frequency and substrate dielectric constant in Figs.~\ref{8}($c$) and ($d$). In these figures, the BA shift \(\Delta\theta^0_{B}\) is more significant at low \(\varepsilon_2\) values and low frequencies. In contrast, the difference \(\delta\theta_B\) is maximized at intermediate frequencies and low \(\varepsilon_2\) values. It is worth noting that the analysis can be generalized by considering a dielectric function that incorporates the phonon modes of the substrates. In this case, the dielectric function would depend on both the frequency and phonon dispersions.

\section{Conclusion}\label{sec:Con}
In conclusion, we have systematically investigated the shifts in Brewster's angle for TM-polarized light in a doped graphene structure, both with and without the influence of the Fizeau drag effect. Our analysis reveals a significant enhancement in the Brewster's angle shift when the Fizeau drag is present, with the shift increasing by several degrees compared to the drag-free scenario. This enhancement introduces a tunable parameter, allowing for precise control over the difference between the Brewster's angle shifts in drag-free and drag-affected cases.

Furthermore, we observed that the Brewster's angle shift can be further amplified by increasing the drift velocity of electrons within the graphene layer. This tunability is complemented by the observation that doping the graphene to increase the number of charged particles also contributes to a larger Brewster's angle shift. These findings highlight the potential for manipulating optical properties in graphene through controlled doping and electron drift, offering promising avenues for the development of advanced optoelectronic devices.

Overall, our results underscore the importance of considering relativistic effects, such as the Fizeau drag, in the design of doped graphene-based optical systems. This work paves the way for future research into the practical applications of graphene in photonic devices, where precise control over light-matter interactions is essential for optimizing device performance.

\begin{acknowledgments}
We acknowledge the financial support from the postdoctoral research grant (ZC304023922) and the NSFC under the grant 
No. 12174346. R. A. received partial funding from the Iran National Science Foundation (INSF) under project No. 4026871.
\end{acknowledgments}

\appendix

\section{Derivation of Eq. (\ref{quartic})}
Substituting the values of $k^i_z$ and $k^t_z$ in Eq. (\ref{BACondition}), it becomes
\begin{equation}
 \frac{\varepsilon_2}{n_1\sqrt{n^2-\sin^2\theta}}-\frac{\varepsilon_1}{n_1\cos\theta}+\sigma=0.\label{A1}
\end{equation}
Dividing both sides by $n_1\sqrt{n^2-\sin^2\theta}.\cos\theta$, the above expression yields
\begin{equation}
{\varepsilon_2}\cos\theta-{\varepsilon_1}\sqrt{n^2-\sin^2\theta}+\sigma n_1\sqrt{n^2-\sin^2\theta}.\cos\theta=0.\label{A2}
\end{equation}
Solving this expression for $\sigma$, we have
\begin{equation}\label{A3}
\sigma=\frac{\varepsilon_1}{n_1\cos\theta}-\frac{\varepsilon_2}{n_1\sqrt{n^2-\sin^2\theta}}.
\end{equation}
Using the fact that $\varepsilon_i=n^2_i$ and rearranging,
\begin{equation}\label{A33}
\sigma=\frac{n_1}{\cos\theta}-\frac{n_1n^2}{\sqrt{n^2-\sin^2\theta}}.
\end{equation}
Let $\cos{\theta}=x$, the above expression takes the form
\begin{equation}\label{A4}
\sigma=n_1\left(\frac{1}{x}-\frac{n^2}{\sqrt{n^2-\sin^2\theta}}\right).
\end{equation}
Using the substitution $n^2=\eta+1$ given in the main text, the above equation further reduces to
\begin{equation}\label{A5}
\frac{\sigma}{n_1}=\frac{1}{x}-\frac{\eta+1}{\eta+x^2},
\end{equation}
where the identity $\sin^2\theta+\cos^2\theta=1$ has been employed. Upon rearranging, squaring both sides and substituting $\sigma'=\sigma/n_1$, we obtain
\begin{equation}\label{A6}
\sigma'^2+\frac{1}{x^2}-\frac{2\sigma'}{x}=\frac{\eta^2+1+2\eta}{\eta+x^2}.
\end{equation}
Multiplying both sides of Eq. (\ref{A6}) by $x^2(\eta+x^2)$ and dividing by $\eta$, we get
\begin{eqnarray}
\frac{\sigma'^2}{\eta}x^4-\frac{2\sigma'}{\eta}x^3+[\sigma'^2-(\eta+2)]x^2-2\sigma'x+1=0
\end{eqnarray}
Replacing $\eta=n^2-1$ in the third term of the above equation, it takes the form given in Eq. (\ref{quartic}) of the main text.
\nocite{*}

\bibliographystyle{apsrev4-2.bst}
\bibliography{apssamp}

\end{document}